\documentclass[a4paper,11pt]{article}
\pdfoutput=1 % if your are submitting a pdflatex (i.e. if you have
\usepackage{jcappub} % for details on the use of the package, please
\usepackage[T1]{fontenc} % if needed

\usepackage{graphicx}% Include figure files
\usepackage{epsfig}
\usepackage{dcolumn}% Align table columns on decimal point
\usepackage{bm}% bold math
\usepackage[utf8]{inputenc}
\inputencoding{utf8}
\usepackage{hyperref}
\usepackage{multirow}
\usepackage{caption}
\usepackage{subcaption}
\usepackage{amssymb}
\usepackage{amsmath}
\usepackage{indentfirst}

\title{\boldmath Inflation in non-minimal matter-curvature coupling theories}

\author[a]{C. Gomes,}
\author[b,1]{J. G. Rosa,\note{Also at Departamento de F\'{\i}sica e Astronomia, Faculdade de Ci\^encias da Universidade do Porto, Rua do Campo Alegre s/n, 4169-007 Porto, Portugal.}}
\author[a]{O. Bertolami}

\affiliation[a]{Departamento de F\'isica e Astronomia and Centro de Física do Porto, Faculdade de Ci\^encias da Universidade do Porto, \\Rua do Campo Alegre s/n, 4169-007 Porto, Portugal}
\affiliation[b]{Departamento de F\'isica da Universidade de Aveiro and CIDMA,\\ Campus de Santiago, 3810-183 Aveiro, Portugal}

\emailAdd{claudio.gomes@fc.up.pt}
\emailAdd{joao.rosa@ua.pt}
\emailAdd{orfeu.bertolami@fc.up.pt}

\abstract{We study inflationary scenarios driven by a scalar field in the presence of a non-minimal coupling between matter and curvature. We show that the Friedmann equation can be significantly modified when the energy density during inflation exceeds a critical value determined by the non-minimal coupling, which in turn may considerably modify the spectrum of primordial perturbations and the inflationary dynamics. In particular, we show that these models are characterised by a consistency relation between the tensor-to-scalar ratio and the tensor spectral index that can differ significantly from the predictions of general relativity. We also give examples of observational predictions for some of the most commonly considered potentials and use the results of the Planck collaboration to set limits on the scale of the non-minimal coupling.}

\begin{document}
\maketitle
\flushbottom

\section{Introduction}
\label{sec:intro}

Inflation is the best-known solution to the initial condition problems of the standard Hot Big Bang model. An inflationary epoch can arise from a modified type of gravity, that includes a quadratic term $ \alpha R^2$ \cite{starobinsky}, or from the dynamics of a scalar field \cite{guth, steinhardt, linde}, the inflaton. Features of the scalar field potential are now quite constrained by the recent measurements of the Cosmic Microwave Background (CMB) \cite{wmap,planck}.

Most of the inflationary models are based on General Relativity (GR). Despite its impressive agreement with astrophysical observations in the solar system, GR requires the introduction of two major dark components in order to account for the cosmological data \cite{GR0, GR}. Alternatively, one may think that GR is not the ultimate theory and several modified gravity models have been developed. Perhaps the simplest ones are the so-called $f(R)$ theories \cite{fr1, fr2, fr3}, in which one replaces the Ricci scalar, $R$, in the Einstein-Hilbert action by a generic function of the scalar curvature. Another interesting possibility is to introduce additionally a non-minimal coupling between matter and curvature, expressed as a product of another function of the curvature $f_2(R)$ and the matter Lagrangian \cite{nmc}. This model has quite interesting implications for stellar stability \cite{stellar}, preheating after inflation \cite{preheating}, as well as in mimicking the dark matter in galaxies and clusters \cite{DM mim1, DM mim2} or the late time accelerated expansion of the Universe \cite{DE mim1}. This scenario has also a non-negligible impact in black hole physics \cite{bh}, and on the generalisation of the Layzer-Irvine equation \cite{linmc}.

The idea that non-minimal couplings can affect inflation is not new. For example, in the Bezrukov-Shaposhnikov model, the Higgs boson acts as the inflaton non-minimally coupled to the scalar curvature \cite{shaposhnikov}, although in this case only a particular function of the field, and not the full scalar Lagrangian, is non-minimally coupled to $R$. 

In this work, we consider the modifications to inflationary scenarios driven by a slowly rolling scalar field introduced by a non-minimal coupling between the inflaton Lagrangian and the Ricci scalar, and show that both the dynamics of inflation and its observational predictions can be considerably different from the minimally coupled case.

This paper is organised as follows. First, we introduce the non-minimal coupling between matter and curvature and discuss the resulting field equations and their main features. We then study the inflationary regime in these theories using the slow-roll approximation. Some potentials for the inflaton are studied and compared to the observational limits set by the Planck collaborations in sections 3 and 4. Our main conclusions are summarised and discussed in section 5.

\section{The non-minimal matter-curvature coupling model}

In the extended $f(R)$ theories with a non-minimal coupling between a curvature term and the matter Lagrangian density, the action functional reads \cite{nmc}:
\begin{equation}
\label{modelo}
S=\int d^4x \sqrt{-g} \left[\kappa f_1\left(R\right) + f_2 \left(R\right)\mathcal{L}\right]~,
\end{equation}
where $f_1(R), f_2(R)$ are arbitrary functions of the Ricci scalar $R$, $\kappa=c^4/(16\pi G)$ and $\mathcal{L}$ is the matter Lagrangian density.

Varying the action with respect to the metric, $g_ {\mu\nu}$, leads to the following field equations:
\begin{equation}
\label{fieldequations}
\left( F_1 + \frac{F_2\mathcal{L}}{\kappa}\right) G_{\mu\nu} = \frac{1}{2\kappa}f_2 T_{\mu\nu} + \Delta_{\mu\nu}\left( F_1 + \frac{F_2\mathcal{L}}{\kappa}\right) + \frac{1}{2}g_{\mu\nu}\left(f_1-F_1R-\frac{F_2R\mathcal{L}}{\kappa}\right) ~,
\end{equation}
where $G_{\mu\nu}$ is the Einstein tensor, $F_i \equiv df_i/dR$ and $\Delta_{\mu\nu}\equiv \nabla_{\mu}\nabla_{\nu} - g_{\mu\nu} \square $. It is straightforward to see that choosing $f_1(R)=R$ and $f_2(R)=1$, one recovers General Relativity.

Using the Bianchi identities in the above field equations, one finds that the matter energy-momentum tensor is not covariantly conserved in this model:
\begin{equation}
\label{nonconservation}
\nabla_{\mu}T^{\mu\nu} = \frac{F_2}{f_2} \left(g^{\mu\nu}\mathcal{L}-T^{\mu\nu}\right) \nabla_{\mu}R ~.
\end{equation}

For a perfect fluid, one can find an extra force acting on a test particle, given by \cite{nmc}:
\begin{equation}
\label{extraforce}
f^{\mu} = \frac{1}{\rho+p} \left[\frac{F_2}{f_2}\left(\mathcal{L}-p\right)\nabla_{\nu}R + \nabla_{\nu}p\right]h^{\mu\nu} ~,
\end{equation}
where $h^{\mu\nu} = g^{\mu\nu} + u^{\mu}u^{\nu}$ is the projection operator and $u^\mu$ denotes the particle's 4-velocity. In these theories, the degeneracy between the Lagrangian choices $\mathcal{L}=-\rho$ or $\mathcal{L}=p$ is lifted in opposition to what happens in GR \cite{lagrangian choices GR}, since it yields different behaviours for the extra force term (See Ref. \cite{lagrangian choices} for a thorough discussion).

%%%%%%%%%%%%%%%%%%%%%%%%%%%%%%%%%%%%%%%%%%%%%%%%%%%%%%%%%%%%%%%%%%%%%%%%%%%%%%%%%%%%%%%%
%%%%%%%%%%%%%%%%%%%%%%%%%%%%%%%%%%%%%%%%%%%%%%%%%%%%%%%%%%%%%%%%%%%%%%%%%%%%%%%%%%%%%%%%
%%%%%%%%%%%%%%%%%%%%%%%%%%%%%%%%%%%%%%%%%%%%%%%%%%%%%%%%%%%%%%%%%%%%%%%%%%%%%%%%%%%%%%%%

\section{Inflation in the NMC model}

We start by assuming the Robertson-Walker metric for a homogeneous and isotropic universe:
\begin{equation}
\label{metric}
ds^2 = -dt^2 + a^2\left(t\right)\gamma_{ij}dx^idx^j~,
\end{equation}
where $\gamma_{ij}$ denotes the comoving spatial metric. Since inflation will exponentially dilute the effects of spatial curvature, thus addressing the flatness problem, we consider for simplicity a flat Euclidean metric $\gamma_{ij}=\delta_{ij}$.

The matter content in the Universe is well approximated by a set of perfect fluids that dominate the energy balance at different epochs. For each fluid, the energy-momentum tensor takes the form:
\begin{equation}
T^{\mu\nu} = \left(\rho+p\right)u^{\mu}u^{\nu} + pg^{\mu\nu} ~,
\end{equation}
where $u^{\mu}$ is the fluid's 4-velocity. The time component of the non-conservation of the stress tensor, $\nabla_{\mu}T^{\mu 0}$, gives
\begin{equation}
\dot{\rho}+3H\left(\rho+p\right) = 
\begin{cases}
	0 & \text{if } \mathcal{L}=-\rho \\
	-\frac{F_2}{f_2}\left(\rho+p\right)\dot{R} & \text{if } \mathcal{L} = p 
\end{cases}~,
\end{equation}
where the dots denote time derivatives.

In the case of a homogeneous scalar field with Lagrangian density
\begin{equation}
\mathcal{L}_\phi=-\frac{1}{2}\partial_{\mu}\phi \partial^{\mu}\phi - V\left(\phi\right)~,
\end{equation}
we have $\rho_\phi = \dot\phi^2/2 + V(\phi)$ and $p_\phi= \dot\phi^2/2 - V(\phi)= \mathcal{L}_\phi$, and the inflaton's dynamics is given by the field equation:
\begin{equation}
\ddot{\phi}+3H\dot{\phi} + V'\left(\phi\right)= -\frac{F_2}{f_2}\dot{R}\dot{\phi} \equiv -\Gamma\dot\phi~,
\end{equation}
where $V'\left(\phi\right)=dV/d\phi$. We thus see that the non-minimal matter-curvature coupling will induce a friction term in the inflaton's equation of motion, likewise the case of warm inflation \cite{warm1, warm2}.

An analogous term also appears in the continuity equation for radiation, such that for  $\mathcal{L}_r=p$, one gets:
\begin{equation}
\dot{\rho}_r + \frac{4}{3}\left(3H+\frac{F_2}{f_2}\dot{R}\right)\rho_r = 0~.
\end{equation}
Integrating the above expression, one obtains \cite{preheating}:

\begin{equation}
\rho_r(t)=\rho_{ri}\left(\frac{a_i}{a(t)}\right)^4 \left[\frac{f_2(R_i)}{f_2(R)}\right]^{4/3}~.
\end{equation}

For the case $\Gamma \equiv \frac{F_2}{f_2}\dot{R} > 0$, we may expect this friction term to damp the inflaton's motion and thus facilitate slow-roll. However, as opposed to the case of warm inflation, $\Gamma>0$ yields a sink rather than a source for the radiation energy density, so that the non-minimal coupling would dilute the latter faster than expansion instead of counter-acting this dilution effect. Thus, only for $\Gamma<0$ may the non-minimal coupling sustain a thermal bath during inflation. As we will see in the next sections, we are interested in modifications of gravity in the strong curvature/density regime, i.e.~in scenarios where $F_2>0$. In addition, during inflation $R\propto H^2$ is a decreasing function of time, so that $\Gamma<0$. However, since $H$ does not vary considerably during inflation, $\dot{R}\propto \epsilon$, where $\epsilon$ is the usual slow-roll parameter (defined below). This implies that the friction term will have a negligible effect on the inflaton's motion and will not be able to counteract the dilution of the radiation energy density. Inflation will thus occur in a supercooled regime unless interactions between the inflaton and other degrees of freedom are additionally considered, as e.g.~in \cite{Bastero-Gil:2016qru}. In this work, we will henceforth neglect the effects of $\Gamma$.

%%%%%%%%%%%%%%%%%%%%%%%%%%%%%%%%%%%%%%%%%%%%%%%%%%%%%%%%%%%%%%%%%%%%%%%%%%%%%%%%%%%%%%%%

\subsection{The slow-roll approximation}

The modified Friedmann equation in the theories under study is written as \cite{modified friedmann}:
\begin{equation}
\label{friedmann}
H^2 = \frac{1}{6}\frac{1}{F_1+\frac{F_2\mathcal{L}}{\kappa}} \left[\frac{f_2\rho}{\kappa}-6H\partial_t\left(F_1+\frac{F_2\mathcal{L}}{\kappa}\right)+\left(F_1+\frac{F_2\mathcal{L}}{\kappa}\right)R-f_1\right] ~.
\end{equation}
From the $00$ and the $ii$ components of the modified field equations, we get \cite{frazao}:
\begin{align}
f_2 \rho &= 3FH^2 + \frac{1}{2}\left(M^2_Pf_1-FR\right) + 3H\dot{F}~, \\
f_2 p &= -3FH^2 - \frac{1}{2}\left(M^2_Pf_1-FR\right) - 2H\dot{F} - \ddot{F} -2\dot{H}F ~,
\end{align}
where $F=M_P^2F_1-F_2\rho$ and $M_P^{-2}=8\pi G$ is the reduced Planck mass. We thus see that $p=-\rho$ up to terms that depend on time derivatives of the scalar curvature $R$ and matter density $\rho$, which vanish in the exact de Sitter limit. Defining the slow-roll parameters:
\begin{equation}
\epsilon = -\frac{\dot{H}}{H^2}~, \qquad \eta = -\frac{\ddot{H}}{2\dot{H}H}~,
\end{equation}
and noting that $R=6\left(\dot{H}+2H^2\right)$, such that
\begin{equation}
R=6H^2\left(2-\epsilon\right) \implies \dot{R}\approx -24H^3 \epsilon~,
\end{equation}
we conclude that a slow-roll regime where one can consistently neglect the time variation of $\rho$ and $H$ requires, as in GR, that the conditions $\epsilon, |\eta|\ll 1$ are satisfied.

In this work we will focus on theories such that the pure gravitational sector of the action has the Einstein-Hilbert form, i.e. $f_1(R)=R$, so as to isolate the effects of the non-minimal coupling between matter and curvature. Taking the slow-roll limit of the modified Friedmann equation (\ref{friedmann}), for $\mathcal{L}=p \simeq-\rho$, we then obtain:
\begin{equation}
\label{modifiedfriedmann}
H^2 \simeq \left(\frac{f_2}{1+\frac{2F_2\rho}{M_P^2}}\right) \frac{\rho}{3M_P^2}~.
\end{equation}

It is thus clear that the Friedmann equation may have a different form depending on the inflaton energy density and on the form of the non-minimal matter-curvature coupling. In particular, we expect a different behaviour in the high and low densities regimes, as we will see in the next sections once we specify the form of $f_2(R)$ \footnote{See Annex.}.

In addition, we note that the $ij$-component of the field equations, for flat FRW model for $f_1(R)=R$ and $\mathcal{L}=p$ is given by:
\begin{equation}
-\left[ 2\frac{\ddot{a}}{a}+H^2\right]\left(1+2\frac{F_2p}{M_P^2}\right) = \frac{f_2p}{M^2_P} + \frac{2}{M_P^2}\frac{\Delta_{ij}\left(F_2p\right)}{g_{ij}}-\frac{1}{M_P^2}F_2pR~.
\end{equation}
As seen above, for this cosmological model, $R=12H^2+6\dot{H}=6H^2+6\frac{\ddot{a}}{a}$ and thus:
\begin{equation}
-\left[ 2\frac{\ddot{a}}{a}+H^2\right]\left(1-\frac{F_2p}{M_P^2}\right) + 3H^2\frac{F_2p}{M_P^2} = \frac{f_2p}{M^2_P}+\frac{2}{M^2_Pa^2}\Delta_{ij}\left(F_2p\right)~.
\end{equation}
Using the Friedmann equation in the slow-roll regime, Eq. (\ref{modifiedfriedmann}), and consistently neglecting $\Delta_{ij}(F_2p)$, one obtains:
\begin{equation}
2\frac{\ddot{a}}{a} = \frac{f_2 \bar{\rho}}{1+F_2\bar{\rho}} \left[1-\frac{1}{3}\left(\frac{1+4F_2\bar{\rho}}{1+2F_2\bar{\rho}}\right)\right] = 2H^2~,
\end{equation}
with $\bar{\rho}=\rho/M^2_P$. Hence, as in GR, we obtain accelerated expansion in the slow-roll regime with $p\simeq-\rho$.

%%%%%%%%%%%%%%%%%%%%%%%%%%%%%%%%%%%%%%%%%%%%%%%%%%%%%%%%%%%%%%%%%%%%%%%%%%%%%%%%%%%%%%%%
%%%%%%%%%%%%%%%%%%%%%%%%%%%%%%%%%%%%%%%%%%%%%%%%%%%%%%%%%%%%%%%%%%%%%%%%%%%%%%%%%%%%%%%%
%%%%%%%%%%%%%%%%%%%%%%%%%%%%%%%%%%%%%%%%%%%%%%%%%%%%%%%%%%%%%%%%%%%%%%%%%%%%%%%%%%%%%%%%

\subsection{Quantum fluctuations}

To determine the spectrum of quantum fluctuations of the non-minimally coupled inflaton field, we must consider the full Klein-Gordon equation, although we consistently neglect the friction term in the slow-roll approximation as explained above:
\begin{equation}
\square\phi = - V'(\phi) ~,
\end{equation}
where $\square$ denotes the D'Alembertian operator.  We may then consider small quantum fluctuations around the classical background value of the scalar field:
\begin{equation}
\phi = \bar{\phi}+\delta \phi~.
\end{equation}

In the slow-roll regime, we may consistently neglect the effective scalar field mass, i.e. $V''(\bar{\phi})$, such that we obtain:
\begin{equation}
\delta\ddot{\phi} + 3H\delta\dot{\phi} - \frac{1}{a^2}\nabla^2 \delta \phi \simeq 0~.
\end{equation}

This is completely analogous to the equation obeyed by field fluctuations in GR, so that the power spectrum of inflaton perturbations in the Bunch-Davies vacuum is given by the well-known expression:
\begin{equation}
P_{\phi}(k)=\frac{H^2}{2k^3}~,
\end{equation}
although we note that this cannot be written in the same way as in GR in terms of the inflaton potential due to the modifications to the Friedmann equation introduced by the non-minimal matter-curvature coupling. The relation between the inflaton and the gauge-invariant comoving curvature perturbation is also unaffected in the slow-roll regime, so that we have for the  dimensionless power spectrum of inflaton perturbations:
\begin{equation}
\Delta_{\mathcal{R}}^2(k)=\left({H\over 2\pi}\right)^2 \left(\frac{H}{\dot{\phi}}\right)^2~.
\end{equation}

As we have mentioned earlier, we are interested in non-minimal couplings that introduce modifications in the high curvature/density regime and reduce to GR in the low-density limit, in particular functions of the form $f_2(R)=1+(R/M^2)^n$ for $n>1$. Since $R$ is quadratic in curvature and tensor perturbations, the non-minimal coupling will not modify the action for these perturbations at the leading quadratic order, such that the spectrum of tensor perturbations also retains its GR form:
\begin{equation}
\Delta^2_t(k) = \frac{8}{M^2_P} \Delta^2_{\phi}(k) = \frac{2}{\pi^2} \frac{H^2}{M^2_P}~.
\end{equation}
%

%%%%%%%%%%%%%%%%%%%%%%%%%%%%%%%%%%%%%%%%%%%%%%%%%%%%%%%%%%%%%%%%%%%%%%%%%%%%%%%%%%%%%%%%
%%%%%%%%%%%%%%%%%%%%%%%%%%%%%%%%%%%%%%%%%%%%%%%%%%%%%%%%%%%%%%%%%%%%%%%%%%%%%%%%%%%%%%%%
%%%%%%%%%%%%%%%%%%%%%%%%%%%%%%%%%%%%%%%%%%%%%%%%%%%%%%%%%%%%%%%%%%%%%%%%%%%%%%%%%%%%%%%%

\subsection{Inflation with a general modified Friedmann equation}

In the previous sections, we have seen that the main modification introduced by the non-minimal coupling in the slow-roll regime is a modified form of the Friedmann equation. This motivates us to discuss in more detail the dynamics of inflation for a generic form $H^2=H^2(\rho)$, which may then be applied to different forms of the non-minimal coupling function $f_2(R)$, as well as in other scenarios where such modifications arise but where the slow-roll form of the inflaton's equation of motion, $3H\dot\phi\simeq -V'(\phi)$, is unaffected.

We have, in general, that:
\begin{equation}
\frac{\ddot{a}}{a}=H^2\left[1-3\frac{dH/d\rho}{H(\rho)}\dot{\phi}^2\right] = H^2 \left[1-\epsilon_H\right]~,
\end{equation}
where we may write the slow-roll parameter in the form $\epsilon = 3\frac{dH/d\rho}{H(\rho)}\dot{\phi}^2$. In order to have $\ddot{a}>0$, we need to satisfy the condition $\epsilon<1$, which we can write in terms of the field slow-roll parameter $\epsilon_\phi= M_P^2(V'/V)^2/2$:
\begin{equation} \label{SR1}
\epsilon_\phi< 3\left({M_P^2 H^2\over V}\right)^2\left({dH^2\over d\bar\rho}\right)^{-1}~,
\end{equation}
for $\rho\simeq V(\phi)$. It is easy to check that this yields $\epsilon_\phi <1$ in the GR limit. We note, in addition, that slow-roll also requires $\rho\simeq -p$, i.e.~$\dot\phi^2/2 <V$, which using the slow-roll equation for the field can be written as:
\begin{equation} \label{SR2}
\epsilon_\phi< 9 {H^2M_P^2\over V}~.
\end{equation}

While in GR the two conditions on the slow-roll parameter $\epsilon_\phi$ are essentially redundant (up to $\mathcal{O}(1)$ factors), in general they yield different constraints, and in determining the end of inflation we must consider the strongest of these conditions.  For example, in the high density regime that we will consider in the next section, $V\gg H^2 M_P^2$, the second condition (\ref{SR2}) yields a stronger constraint than Eq.~(\ref{SR1}).

The number of e-folds of inflationary expansion is given, in general, by:
\begin{equation}
N_e = \int H dt = \int \frac{H}{\dot{\phi}} d\phi \approx -3 \int_{\phi_i}^{\phi_e} \frac{H^2(\rho)}{V'}d\phi~.
\end{equation}

We can also determine several quantities which are related to CMB observational data, such as the amplitude of the dimensionless scalar curvature power spectrum:
\begin{equation}
\Delta^2_{\mathcal{R}} = \left(\frac{H}{\dot{\phi}}\right)^2 \left(\frac{H}{2\pi}\right)^2 = \frac{9H^6(\rho)}{4\pi^2V'^2}= {1\over 24\pi^2}{V\over M_P^4}\epsilon_\phi^{-1} \left({3H^2M_P^2\over V}\right)^3 ~,
\end{equation}
where one can clearly see that the amplitude of curvature perturbations is suppressed for $V\gg H^2M_P^2$. The associated scalar spectral index is then given by:
\begin{eqnarray}
n_s-1 &\equiv& \frac{d \ln \Delta^2_{\mathcal{R}}}{d \ln k} = \frac{\dot{\phi}}{H} \frac{d \ln \Delta^2_{\mathcal{R}}}{d\phi} = -2\frac{dH}{d\rho}\frac{V'^2}{H^3} + \frac{2}{3}\frac{V''}{H^2} \nonumber\\
& \simeq & -6 \left(3{dH^2\over d\bar\rho}\right) \left(V\over 3H^2M_P^2\right)^2 \epsilon_\phi +2 \left(V\over 3H^2M_P^2\right) \eta_\phi ~,
\end{eqnarray}
where $\eta_\phi = M_P^2 V''/V$ satisfies the slow-roll condition $|\eta_\phi|\ll H^2 M_P^2 /V $ analogous to Eq.~(\ref{SR2}).

From the amplitude of the tensor spectrum:
\begin{equation}
\Delta^2_t = \frac{8}{M_P^2}\left(\frac{H}{2\pi}\right)^2 = \frac{2}{\pi^2 M^2_P}H^2(\rho)~,
\end{equation}
we may compute the tensor spectral index:
\begin{equation}
n_t \equiv \frac{d \ln \Delta^2_t}{d \ln k} \simeq  - \frac{2V'^2}{3H^3(\rho)} \frac{dH}{d\rho} \simeq  -2 \left(3{dH^2\over d\bar\rho}\right) \left(V\over 3H^2M_P^2\right)^2 \epsilon_\phi~,
\end{equation}
and the tensor-to-scalar ratio
\begin{equation}
r \equiv \frac{\Delta_t^2}{\Delta_s^2} =\frac{8}{9} \frac{V'^2}{M^2_P}\frac{1}{H^4(\rho)} \simeq 16  \left(V\over 3H^2M_P^2\right)^2 \epsilon_\phi~.
\end{equation}
Some of these quantities have been measured or constrained by the Planck collaboration \cite{planck}:
\begin{equation}
n_s = 0.9603 \quad , \qquad
A_s = \Delta^2_{\mathcal{R}} (k=k_*) = 2.2 \times 10^{-9} \quad  , \qquad
r < 0.11 ~.
\end{equation}

The non-minimal coupling between matter and curvature does not change the Lyth bound obtained in GR, since for any modified Friedmann equation of the form $H^2=H^2(\rho)$ we get:
\begin{equation}
\frac{\Delta \phi}{M_P^2} = \mathcal{O}(1) \sqrt{\frac{r}{0.01}} ~.
\end{equation}

The consistency relation between the tensor-to-scalar ratio and the tensor spectral index is in general given by:
\begin{equation}
\label{rnt}
r=-8 \left(3{dH^2\over d\bar\rho}\right)^{-1} n_t~,
\end{equation}
and as we will discuss in more detail below, this may yield considerably different results from the GR case, possibly yielding a `smoking gun' for a non-minimally coupled inflationary regime.

%%%%%%%%%%%%%%%%%%%%%%%%%%%%%%%%%%%%%%%%%%%%%%%%%%%%%%%%%%%%%%%%%%%%%%%%%%%%%%%%%%%%%%%%
%%%%%%%%%%%%%%%%%%%%%%%%%%%%%%%%%%%%%%%%%%%%%%%%%%%%%%%%%%%%%%%%%%%%%%%%%%%%%%%%%%%%%%%%
%%%%%%%%%%%%%%%%%%%%%%%%%%%%%%%%%%%%%%%%%%%%%%%%%%%%%%%%%%%%%%%%%%%%%%%%%%%%%%%%%%%%%%%%

\subsection{The high density limit}

Let us consider now the high density regime of the inflationary scenario in these alternative theories of gravity.  We shall study some forms for the non-minimal coupling function $f_2(R)$ which have been used in the literature to address issues as dark matter, dark energy, reheating after inflation, and black holes, namely exponential and power-law types \cite{DM mim1, DM mim2, DE mim1,preheating, bh}.

If e.g.~the non-minimal coupling function is chosen to be of the form $f_2(R)=e^{R/M^2}$, the modified Friedmann Eq. (\ref{modifiedfriedmann}) becomes simply $\frac{R}{M^2}\simeq2$ for $\rho/(M_P^2M^2)\gtrsim 0.07$, which means that $H \simeq M/\sqrt{6}$. For $M \ll M_P$, this solution is achieved for energy densities far below the Planck scale.

Another class of functions that may be of interest is the power-law type, $f_2(R)=1+\left(\frac{R}{M^2}\right)^n$, whose asymptotic solution is:
\begin{equation}
\frac{R}{M^2} = \left(\frac{2}{n-2}\right)^{1/n} + \mathcal{O}\left(\frac{M_P^2M^2}{\rho}\right) \quad, \qquad n>2 ~.
\end{equation}

We point out that for $n=2$ the modified Friedmann equation yields $H^2=\rho/3M_P^2$, which is the standard solution in GR that is now retrieved not only for $f_2(R)=1$ (as expected) but also for $f_2(R)=1+(R/M^2)^2$. For $n<2$ there are no real solutions for $R/M^2$, therefore this case will not be considered. For $n>2$, we note that, as for the exponential function considered above, the solution yields $R/M^2 \sim \mathcal{O}(1)$, so that $H \sim M$ up to numerical factors. This implies that this solution is attained for energy densities
\begin{equation}
\rho  \gtrsim M_P^2M^2 ~,
\end{equation}
which are sub-Planckian for $M \ll M_P$. For example, for the case $n=3$ the Friedmann equation yields, for $\rho \gtrsim M_P^2M^2$:
\begin{equation}
H^2 \simeq \frac{M^2}{12} \left(2^{1/3} - \frac{1}{6\times 2^{1/3}}\frac{M_P^2M^2}{\rho}\right) ~.
\end{equation}

\begin{figure} [htbp]
\centering
\includegraphics[scale=0.9]{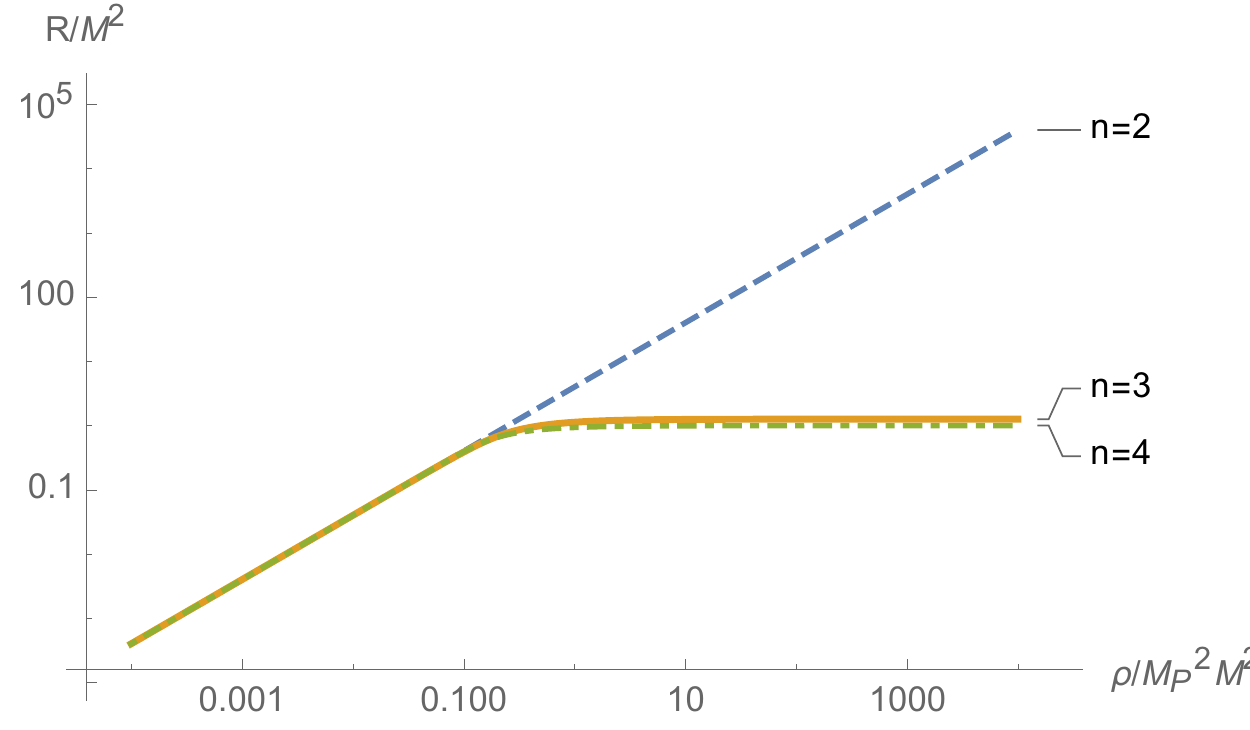}
\caption{Solutions of the Friedmann equation for power-law non-minimal coupling functions, $f_2(R)=1+\left(\frac{R}{M^2}\right)^n$. The case $n=2$ corresponds exactly to the GR solution. For $n>2$ the behaviour is quite similar for different powers.}
\label{fig:friedmann}
\end{figure}

In Fig.~\ref{fig:friedmann} we give examples of the modified Friedmann equation $H^2(\rho)$ for different power-law functions, showing that for $n>2$ the asymptotic regime is quickly achieved, whilst the low density regime corresponds to the standard Friedmann expression in GR. Since there is no significant deviation between power-law functions with indices $n>2$, we shall study the cubic case in the next section  which is the simplest monomial that yields a non-trivial correction to GR in the inflationary regime, pointing out that for other powers the analysis is quite similar.

%%%%%%%%%%%%%%%%%%%%%%%%%%%%%%%%%%%%%%%%%%%%%%%%%%%%%%%%%%%%%%%%%%%%%%%%%%%%%%%%%%%%%%%%
%%%%%%%%%%%%%%%%%%%%%%%%%%%%%%%%%%%%%%%%%%%%%%%%%%%%%%%%%%%%%%%%%%%%%%%%%%%%%%%%%%%%%%%%
%%%%%%%%%%%%%%%%%%%%%%%%%%%%%%%%%%%%%%%%%%%%%%%%%%%%%%%%%%%%%%%%%%%%%%%%%%%%%%%%%%%%%%%%

\section{Cubic non-minimal coupling: $f_2(R)=1+\left(\frac{R}{R_0}\right)^3$}

For $f_2(R)=1+\left(\frac{R}{M^2}\right)^3$, and defining $H_0^2=2^{1/3}M^2/12$ and $\tilde{\rho}=\rho/M_P^2H_0^2$, the modified Friedmann equation yields:
\begin{equation}
H^2=H_0^2 \frac{-2^{2/3} \tilde{\rho} + \left(\tilde{\rho}^3 + \sqrt{\tilde{\rho}^3 (4 + \tilde{\rho}^3)}\right)^{
  2/3}}{2^{1/3} \tilde{\rho} \left(\tilde{\rho}^3 + \sqrt{\tilde{\rho}^3 (4 + \tilde{\rho}^3)}\right)^{
 1/3}} ~.
\end{equation}

Since this is a rather cumbersome function, we use the simpler approximation:
\begin{equation} \label{friedmann_approx}
H^2=H_0^2 \frac{\tilde{\rho} + \tilde{\rho}^2}{3 + 2 \tilde{\rho} + \tilde{\rho}^2}~,
\end{equation}
in order to get faster and still accurate numerical and graphical solutions. This approximation still gives the correct low and high density limits and is very close in the intermediate density region.

The scalar index and the tensor-to-scalar ratio read:
\begin{eqnarray}
n_s &=& 1- 2\epsilon_{\phi} \left(\frac{\tilde{\rho}^2+6\tilde{\rho}+3}{\tilde{\rho}^2+2\tilde{\rho}+1}\right) +\frac{2}{3} \eta_{\phi}\left(\frac{\tilde{\rho}^2+2\tilde{\rho}+3}{\tilde{\rho}+1}\right) ~,\\
r&=&\frac{16}{9}\epsilon_{\phi} \left(\frac{\tilde{\rho}^2+2\tilde{\rho}+3}{\tilde{\rho}+1}\right)^2 ~.
\end{eqnarray}

It is easy to check that in the low-density limit, $\tilde{\rho} \ll 1$, we recover the GR results $n_s-1=-6\epsilon_{\phi}+2\eta_{\phi}$ and $r=16\epsilon_\phi$, while at high densities, $\tilde{\rho} \gg 1$, we obtain
\begin{eqnarray}
n_s &\simeq& 1- 2\epsilon_{\phi}  +\frac{2}{3} \eta_{\phi}\frac{V}{H^2_0 M_P^2}~,\\
r&\simeq&\frac{16}{9}\epsilon_{\phi} \left(\frac{V}{H^2_0M_P^2}\right)^2 ~.
\end{eqnarray}

Below we compare the observational predictions in the $n_s-r$ plane for different forms of the scalar potential in the cubic non-minimal coupling scenario with the constraints obtained by the Planck collaboration \cite{planck}. For all cases we have developed a numerical code to compute the number of e-folds using the approximate form of the Friedmann equation in (\ref{friedmann_approx}) and determining the end of inflation via the strongest of the conditions (\ref{SR1}) and (\ref{SR2}). Fixing the number of e-folds to be between 50 and 60 in order to solve the flatness and horizon problems, we thus determine the field value at which the CMB fluctuations at the pivot scale become super-horizon, and from this we can compute $n_s$ and $r$ as a function of the potential parameters. 

For each potential, we define the dimensionless parameter $x\equiv V_0/M_P^2H_0^2$, where the constant $V_0$ sets the scale of the inflationary potential, such that for $x\ll1$ we recover the GR limit. For potentials characterised by more than one parameter, such as the hilltop and the Higgs-like potentials, we show the results for a given value of $x$ and discuss how the observational predictions change as we vary $\gamma$. For the monomial models, defined uniquely in terms of $V_0$, we show the results as a function of $x$.

%%%%%%%%%%%%%%%%%%%%%%%%%%%%%%%%%%%%%%%%%%%%%%%%%%%%%%%%%%%%%%%%%%%%%%%%%%%%%%%%%%%%%%%%
%%%%%%%%%%%%%%%%%%%%%%%%%%%%%%%%%%%%%%%%%%%%%%%%%%%%%%%%%%%%%%%%%%%%%%%%%%%%%%%%%%%%%%%%

\subsection{Hilltop Potentials: $V=V_0 \left[ 1-\frac{\gamma}{n}\left(\frac{\phi}{M_P}\right)^n \right] $}

\begin{figure}[htbp]
\centering
\includegraphics[scale=0.6]{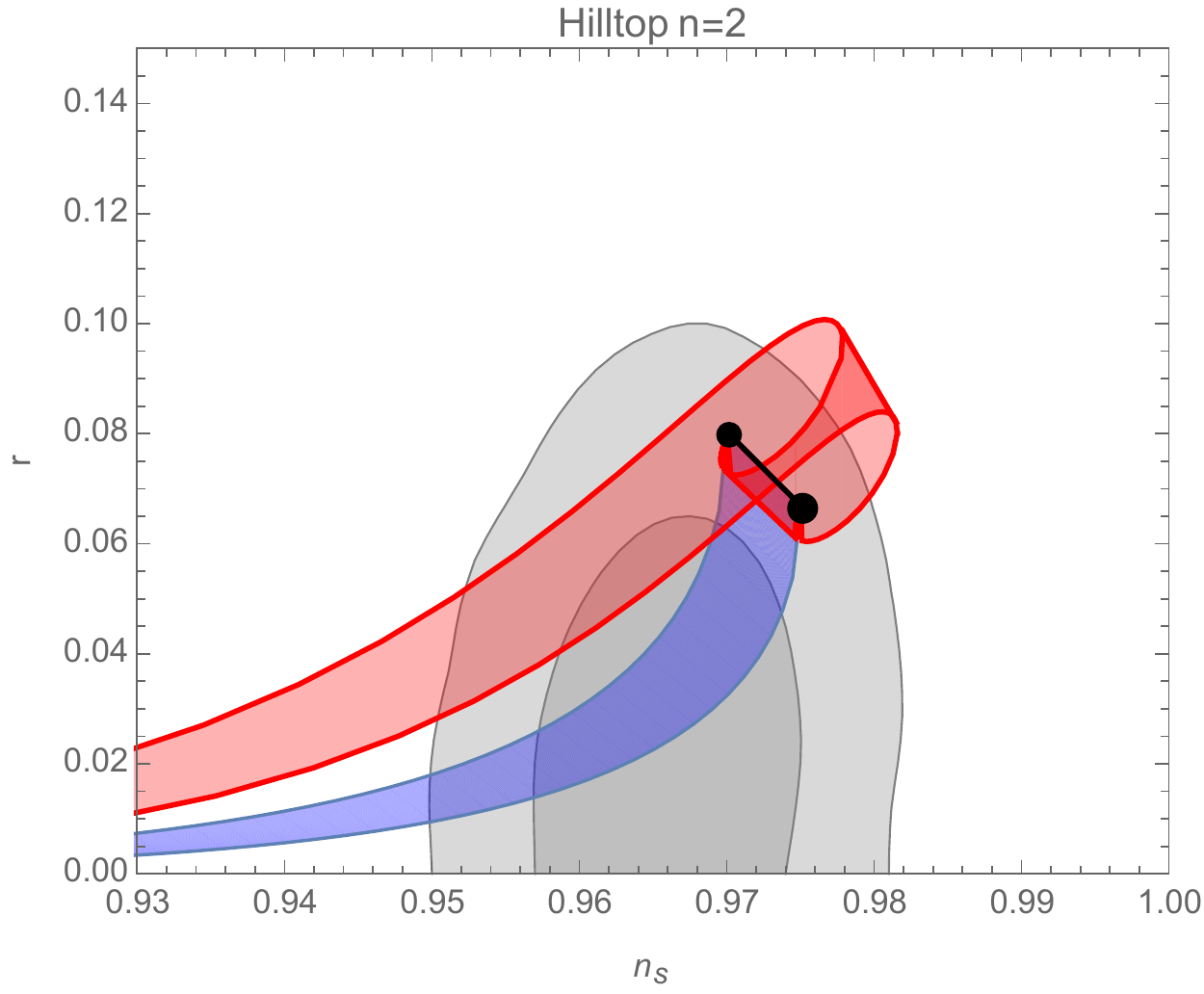}
\includegraphics[scale=0.6]{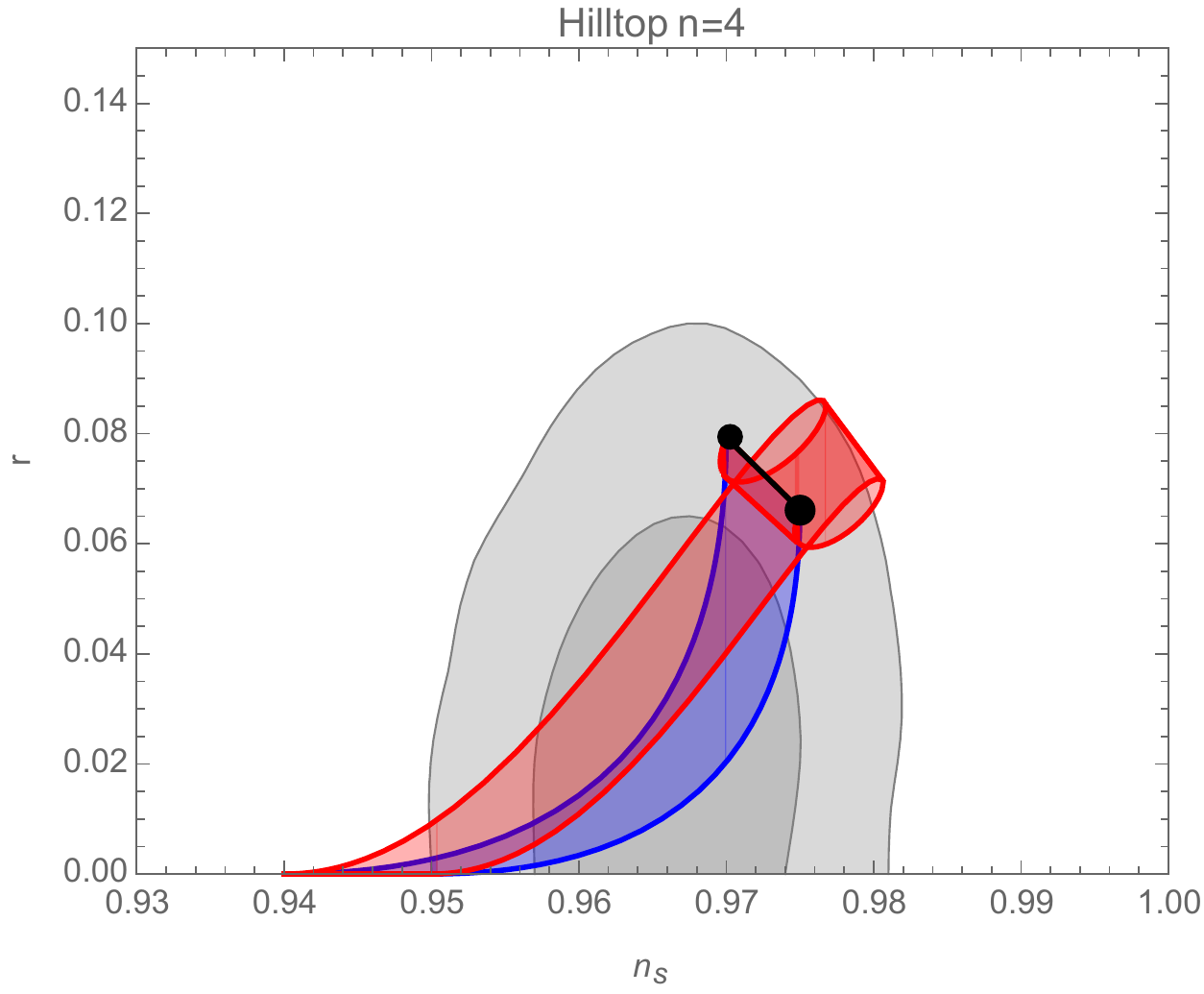}
\caption{Observational predictions for the quadratic (left) and quartic (right) hilltop potentials in the NMC model with $x=5$ are shown in red, and the corresponding GR predictions are shown in blue. Upper and lower bounds correspond to $N_e=50$ and $N_e=60$, respectively. The black circles correspond to the GR prediction for a linear potential. The light and dark gray regions correspond, respectively, to the 68\% and 95\% C.L. contours obtained by the Planck collaboration \cite{planck}.}
\label{hilltop2}
\end{figure}

In Fig.~\ref{hilltop2} we show the observational predictions for the quadratic and quartic hilltop potentials, $n=2$ and $n=4$, respectively, setting $x=5$, as a function of the parameter $\gamma$.

For these potentials, the non-minimal coupling generically increases the value of the tensor-to-scalar ratio, which results in slightly higher values for the tensor modes than in GR, particularly for the quadratic model. In both plots, $\gamma$ decreases from left to right and there is a turning point where the potential starts to leave the high density regime, which occurs for smaller values of $\gamma$, and then enters the low density regime. This means that for $\gamma\to 0$ we recover the GR limit in both cases, corresponding to the prediction for a linear scalar potential $V(\phi)\propto \phi$. We also recover the GR limit for $x\rightarrow 0$ for any value of $\gamma$.

For a number of e-folds between 50 and 60, the hilltop models are always compatible with Planck's data \cite{planck} at the $2\sigma$ level, regardless the value of the ratio $V_0/M_P^2H_0^2$. This happens for the branch near the GR behaviour, i.e.~for smaller values of $\gamma$.

%%%%%%%%%%%%%%%%%%%%%%%%%%%%%%%%%%%%%%%%%%%%%%%%%%%%%%%%%%%%%%%%%%%%%%%%%%%%%%%%%%%%%%%%
%%%%%%%%%%%s%%%%%%%%%%%%%%%%%%%%%%%%%%%%%%%%%%%%%%%%%%%%%%%%%%%%%%%%%%%%%%%%%%%%%%%%%%%%

\subsection{Higgs-like Potential: $V=V_0 \left[ 1-\frac{\gamma}{2}\left(\frac{\phi}{M_P}\right)^2 +\frac{\gamma^2}{16}\left(\frac{\phi}{M_P}\right)^4\right] $}

\begin{figure}[htbp]
\centering
\begin{subfigure}{.48\textwidth}
  \centering
  \includegraphics[width=1\linewidth]{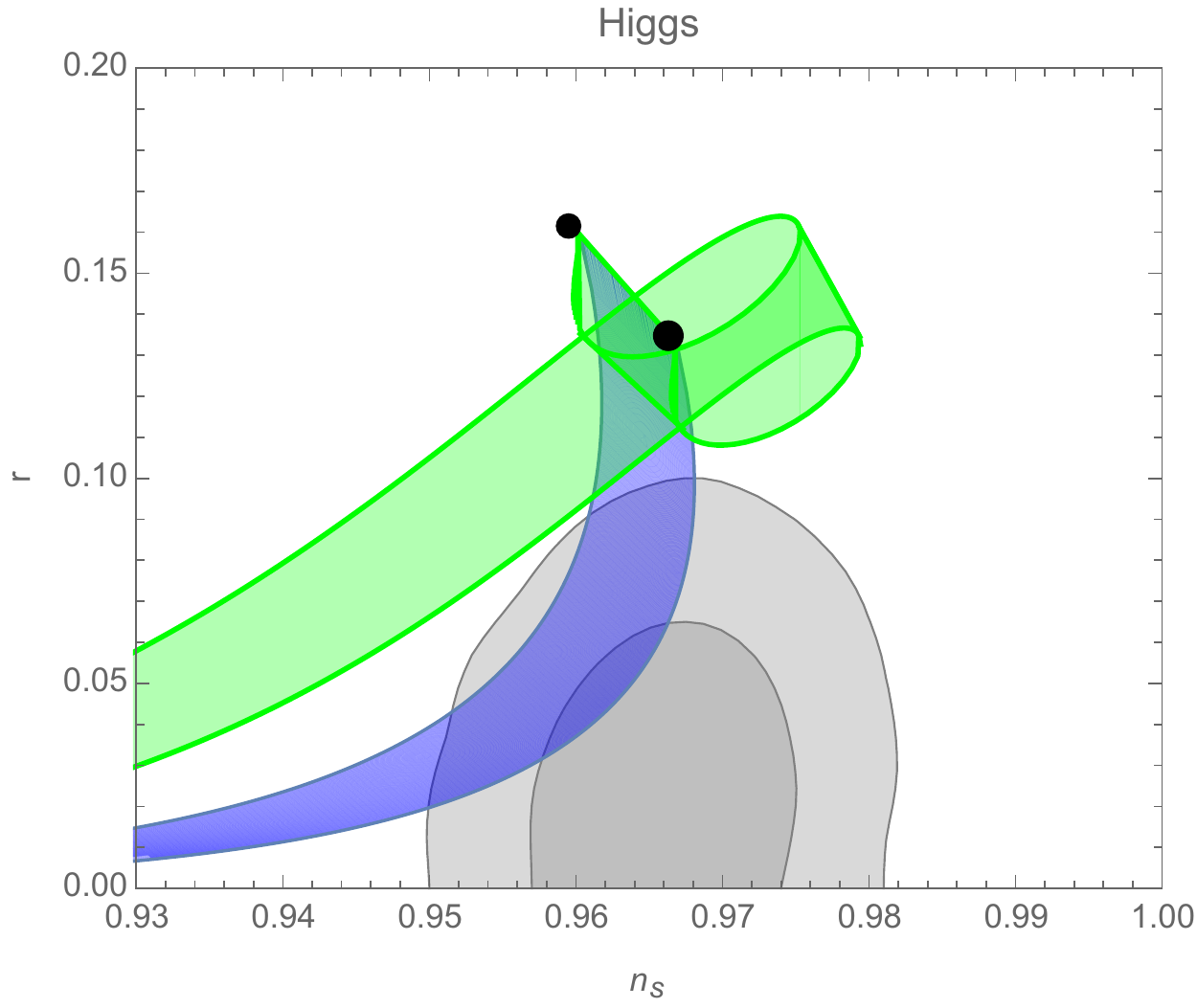}
  \caption{Higgs potential for $x=V_0/M_P^2H_0^2=5$}
  \label{higgs}
\end{subfigure}%
\begin{subfigure}{.48\textwidth}
  \centering
  \includegraphics[width=1\linewidth]{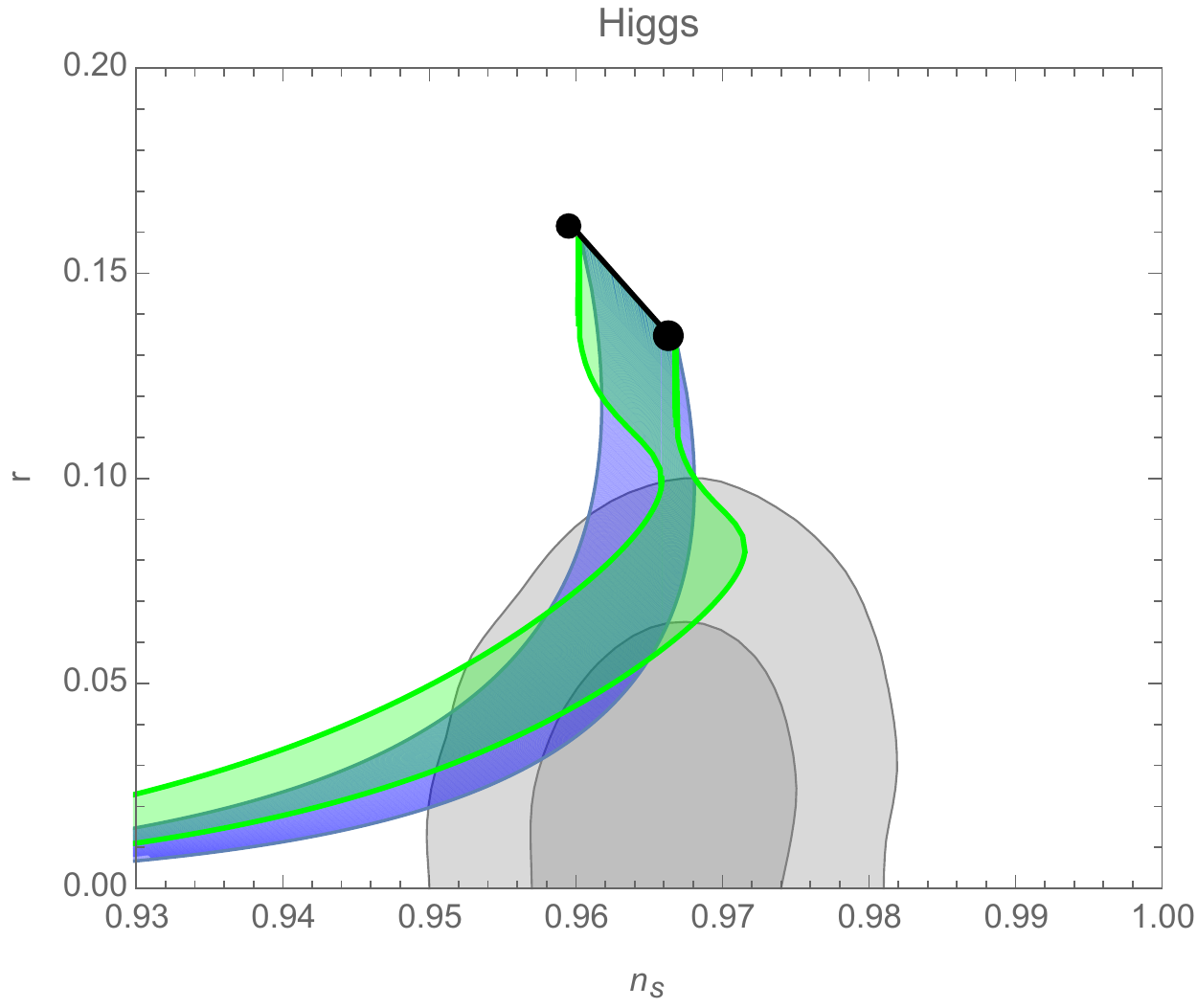}
  \caption{Higgs potential for $x=V_0/M_P^2H_0^2=2$.}
  \label{fig:higgslow}
\end{subfigure}
\caption{Observational predictions for the Higgs-like potential in the NMC model with $x=5$ (left) and $x=2$ (right) are shown in green, and the corresponding GR predictions are shown in blue. Upper and lower bounds correspond to $N_e=50$ and $N_e=60$, respectively. The black circles correspond to the GR prediction for a quadratic potential. The light and dark gray regions correspond, respectively, to the 68\% and 95\% C.L. contours obtained by the Planck collaboration \cite{planck}. The parameter $\gamma$ decreases from nearly 1 to 0 from left to right in both plots.}
\label{fig:HIGGS}
\end{figure}

For a Higgs-like potential, $V(\phi)=\lambda(\phi^2-v^2)^2$, which can be written in the form  $V=V_0 \left[ 1-\frac{\gamma}{2}\left(\frac{\phi}{M_P}\right)^2 +\frac{\gamma^2}{16}\left(\frac{\phi}{M_P}\right)^4\right] $, one proceeds as for the hilltop models above. Note that this potential can be seen as a completion of the quadratic hilltop model that is bounded from below. The observational predictions for $x=5$ and $x=2$ are shown in Fig. \ref{fig:HIGGS}.

In this scenario the value of the tensor-to-scalar ratio is boosted as for the hilltop potentials. For $x=V_0/M_P^2H_0^2 \gtrsim 4.8$, the Higgs model is ruled out in the NMC scenario by the observational constraints imposed by Planck data \cite{planck}, as shown in Fig. \ref{higgs}. For a smaller ratio, the resulting behavior is similar to GR; for instance, if we set that ratio to be $V_0/M_P^2H_0^2=2$, we get the results shown in Fig. \ref{fig:higgslow}. For $x \lesssim 2.6$ the Higgs potential is compatible with observations mostly at the $2\sigma$ level, although there are small parametric regions where agreement is found at $1 \sigma$.

%%%%%%%%%%%%%%%%%%%%%%%%%%%%%%%%%%%%%%%%%%%%%%%%%%%%%%%%%%%%%%%%%%%%%%%%%%%%%%%%%%%%%%%%
%%%%%%%%%%%s%%%%%%%%%%%%%%%%%%%%%%%%%%%%%%%%%%%%%%%%%%%%%%%%%%%%%%%%%%%%%%%%%%%%%%%%%%%%

\subsection{Chaotic models: $V=V_0 \left(\frac{\phi}{M_P}\right)^n $}

In chaotic (monomial) models, observational predictions depend exclusively on the ratio $x=V_0/H_0^2M_P^2$. In general, the potential can be cast in the form:
\begin{equation}
V=V_0 \left(\frac{\phi}{M_P}\right)^n ~.
\end{equation}

\begin{figure}[htbp]
\centering
\includegraphics[scale=0.7]{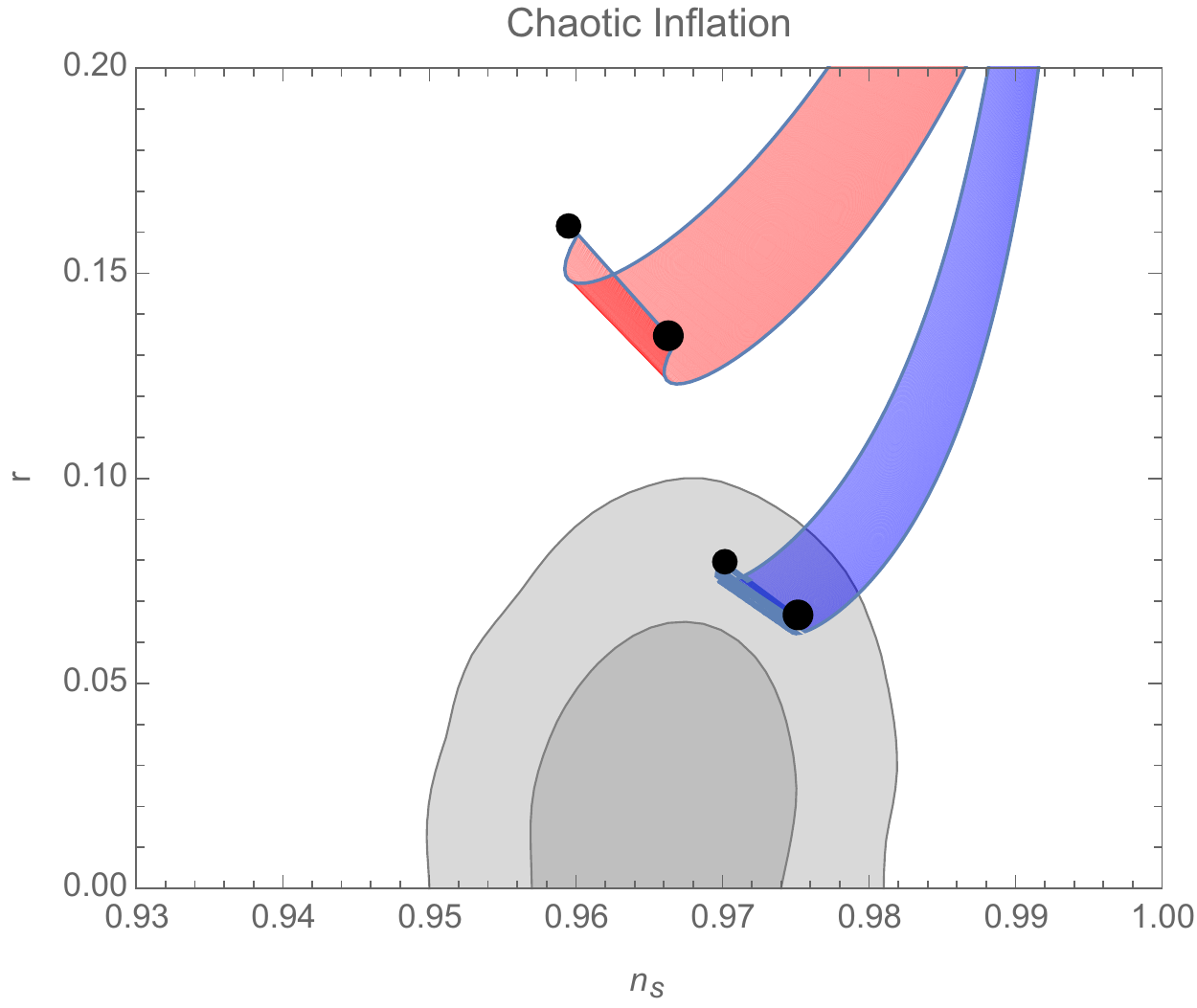}
\caption{Observational predictions for chaotic potentials with $n=1$ shown in blue and $n=2$ shown in red. The parameter $x=V_0/M_P^2H_0^2$ decreases towards zero from top to bottom. Upper and lower bounds correspond to $N_e=50$ and $N_e=60$, respectively. The black circles correspond to the corresponding GR predictions.}
\label{fig:chaotic}
\end{figure}

In GR, only monomial potentials with $n<2$ are consistent with the Planck results \cite{planck}. In the NMC scenario, we find this is also the case, although we have a further constraint on the ratio $x$. For $n=1$ and $n=2$, we find the observational predictions shown in Fig.~\ref{fig:chaotic}, where it is clear that $n=2$ is ruled out, while the linear potential with $n=1$ is allowed at the $2\sigma$ confidence level for $x \lesssim 0.2$. 

%%%%%%%%%%%%%%%%%%%%%%%%%%%%%%%%%%%%%%%%%%%%%%%%%%%%%%%%%%%%%%%%%%%%%%%%%%%%%%%%%%%%%%%%
%%%%%%%%%%%s%%%%%%%%%%%%%%%%%%%%%%%%%%%%%%%%%%%%%%%%%%%%%%%%%%%%%%%%%%%%%%%%%%%%%%%%%%%%

\subsection{Consistency relation}
\label{subsec:consistency}

\begin{figure}[h!]
\centering
\includegraphics[scale=0.7]{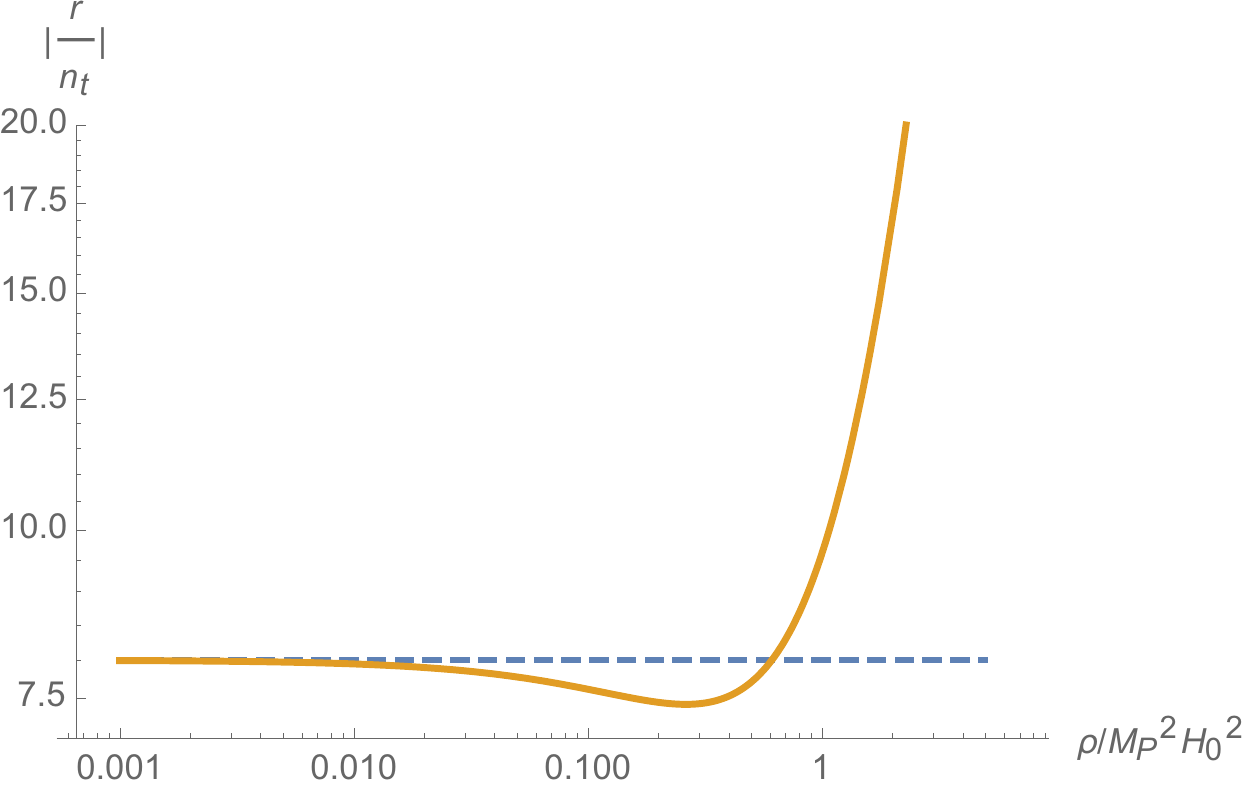}
\caption{Variation of the ratio $|r/n_t|$ in the cubic NMC model as a function of the energy density (solid orange curve), with the GR prediction $r=8|n_t|$ given by the dashed blue line.}
\label{fig:rnt}
\end{figure}

As we have discussed earlier, the non-minimal coupling between the inflaton and curvature modifies the standard consistency relation found in GR between the tensor-to-scalar ratio and the tensor spectral index. In Fig.~\ref{fig:rnt}, we plot the variation of the ratio $|r/n_t|$ in the cubic scenario as a function of the normalised energy density at horizon-crossing.

As one can see, for small energy density, $ \rho\leq 0.01 H_0^2 M_P^2$, we recover the GR limit, as expected since the Friedmann equation reduces to the standard GR form in this limit. For intermediate values, in the range $[0.01,0.6]$, we find a ratio smaller than the GR value. This is the behaviour also obtained in warm inflation \cite{warmcr}, or in multifield models of inflation \cite{multifield0, multifield, multifield2}. 

Energy densities higher than $0.6H_0^2M_P^2$ exhibit a larger $|r/n_t|$ ratio, meaning that for $r \lesssim 0.11$ tensor modes should exhibit a very nearly scale invariant spectrum. This may be also the case in models where the initial inflaton state is not the Bunch-Davies vacuum \cite{nbdcr1,nbdcr2,nbdcr3,nbdcr4,nbdcr5}. Such large deviations in the high density limit imply that future observational constrains in both tensor-to-scalar ratio and the tensor index may provide a tool to clearly distinguish between General Relativity and the NMC matter-curvature models.

In addition, we note that in the high-density regime we have for the amplitude of the scalar power spectrum:
\begin{equation}
A_s \simeq \frac{2}{\pi^2 M_P^2} \frac{H_0^2}{r} ~.
\end{equation}

Since for the models analysed above agreement with observations is found for scenarios where $r\sim 10^{-2}$, we can infer from this the magnitude of the non-minimal coupling mass scale to be $M \sim 10^{-5} M_P \sim 10^{13}$ GeV, which perhaps not surprisingly, is comparable to the magnitude of the Hubble parameter in most inflationary models in GR.

After inflation, as in GR, the inflaton will transfer its energy into other light particle degrees of freedom through the standard reheating process. Since the radiation energy density $\rho_R~ T^4$, for $T< \sqrt{M M_P}$ the effects of the non-minimal coupling function should go away in the scenario that we are considering. For $M\sim10^{13}$ GeV, this means that for temperatures below $10^{15}$ GeV the non-minimal coupling should have no interesting effect.

In a previous paper \cite{preheating}, one important conclusion was drawn: the non-minimal coupling could not produce inflation by itself, but was nevertheless important in the reheating and preheating epochs. In the present case, inflation is driven by a real homogeneous scalar field, and the non-minimal coupling function is important in this scenario, but not afterwards since the curvature/energy density decreases after inflation. Therefore we expect reheating to occur in the same fashion as in the GR framework (see e.g. \cite{reheating}).

\section{Conclusions} 

In this work we have analysed the dynamics and observational predictions of scalar field inflation in the presence of a non-minimal matter-curvature coupling. The latter corresponds to a term $f_2(R)\mathcal{L}$ coupling the matter Lagrangian density, in particular that of the inflaton scalar field, to a generic function of the Ricci curvature scalar. We have shown that this leads to a modified Friedmann equation in the slow-roll regime that can deviate significantly from the corresponding GR equation for large energy densities when $f_2'(R)>0$. The inflationary dynamics in GR is nevertheless recovered in the limit of low energy density, and the transition between the two regimes is set by the threshold mass scale that characterises the non-minimal coupling function. In particular, for $f_2(R)=1 + (R/M^2)^n$, with $n>2$, we have found that the Hubble parameter becomes independent of the energy density in the high-density regime, $H(\rho)\simeq H_0$, within the slow-roll approximation.

As a result of the modified Friedmann equation, the NMC at high density leads to significant departures from GR for inflationary observables such as the scalar and tensor spectral indices and the tensor-to-scalar ratio. The density independence of the Hubble parameter, in particular, leads generically to larger values of the tensor-to-scalar ratio than in GR, which has allowed us to use the latest results by the Planck collaboration to set stringent limits on the dimensionless ratio $x=V_0/M_P^2H_0^2$ for some of the most commonly considered scalar potentials, with $V_0$ setting the inflationary energy scale. There are, however, specific forms of the scalar potential, such as e.g.~hilltop models, where we found parametric regions that are observationally consistent independently of $x$.

In performing this study, we have also derived the form of inflationary observables for a generic modified Friedmann equation $H^2=H^2(\rho)$, and which can be applied to other models where such modifications occur.

Furthermore, we have shown that the $|r/n_t|$ ratio, which constitutes a consistency relation for single field models in GR, may deviate significantly from the latter's prediction, particularly at high densities, where we have found $|r/n_t|\gtrsim 8$. This implies that future observational constraints in both the tensor-to-scalar ratio and the tensor index may help distinguishing GR from the NMC curvature-matter models of gravity.

\section{Avoidance of singularities}
In this appendix we address the issue of singularities  related to the change of sign of the gravitational constant.

In alternative theories of gravity it is common to express the modified field equations in terms of the Einstein tensor and effective energy-momentum tensors. Eq. (\ref{fieldequations}) can be cast in the form:
\begin{equation}
\label{effectiveconstant}
 G_{\mu\nu} = \frac{f_2}{\left( F_1 + \frac{F_2\mathcal{L}}{\kappa}\right)}\frac{1}{2\kappa}\left[ T_{\mu\nu} +\hat{T}_{\mu\nu}\right] ~,
\end{equation}
with the effective energy-momentum tensor defined as
\begin{equation}
f_2\hat{T}_{\mu\nu} \equiv \Delta_{\mu\nu}\left( F_1 + \frac{F_2\mathcal{L}}{\kappa}\right) + \frac{1}{2}g_{\mu\nu}\left(f_1-F_1R-\frac{F_2R\mathcal{L}}{\kappa}\right) ~.
\end{equation}

From Eq.(\ref{effectiveconstant}) one can define an effective gravitational constant according to the similarity with Einstein's field equations: $\hat{G}_{eff}\equiv \frac{f_2}{\left( F_1 + \frac{F_2\mathcal{L}}{\kappa}\right)}\frac{1}{2\kappa}$. Although convenient, this kind of identifications may present singularities, specially if it changes sign and one considers homogeneous but anisotropic models \cite{negativeconstant, anisotropies}. Since inflation needs to work beyond the assumptions of homogeneity and isotropy, this may present a problem in our model. 

Nonetheless, when deriving the Friedmann equation during slow-roll for $f_1=R$ in Eq. (\ref{modifiedfriedmann}), we have obtained a natural effective gravitational constant related to the quantity $G_{eff}\sim \frac{f_2}{1+\frac{2F_2\rho}{M_P^2}}G$ which results from the balance between the convenient constant $\hat{G}_{eff}$ and the effective energy-momentum tensor. This quantity is always positive during slow-roll. And after slow-roll, we need to look at the previous constant which will be positive since the inflaton will be kinetically dominated.

Nevertheless, if a zero occurs in the expression $1-2\frac{F_2V(\phi)}{M_P^2}$ (related to $\hat{G}_{eff}$) and cannot be avoided when considering, for instance, field and metric perturbations, then one may have some restrictions
on the free parameters for each inflationary potential discussed in this paper, $\left(x, \gamma\right)$, imposing that the effective gravitational constant is positive during inflation. This happens for the hypersurface $
\frac{V(\phi)}{V_0}\approx\frac{2.57653}{x}$, for $f_1=R$ and $f_2=1+\left(\frac{R}{M^2}\right)^3$ during slow-roll. However, the regions of those parameters are still compatible with Planck's data. For the case of chaotic $n=1$ model, this requirement implies that $x\lesssim 0.24$ for $50$ e-folds and $x\lesssim 0.22$ for $N_e=60$, which means that $r<0.11$. Although ruled out from observational data, chaotic $n=2$ potential requires $x\lesssim 0.012$ for $N_e=50$ and $x\lesssim 0.01$ for $N_e=60$. For the quadratic hilltop potential, this means for $x=5$ the coupling parameter $\gamma \lesssim 0.0018$ when $N_e=50$ and $\gamma \lesssim 0.0015$ for $N_e=60$, which graphically implies that the region from the turning around point till $\gamma \to 0$ is allowed. However, as we lower the ratio $x$ higher values for the parameter are allowed; for instance, if $x=2$ all values in the range $\gamma \in ]0,1[$ are allowed. The case of the quartic hilltop potential is similar: for $x=5$ the free parameter has to be $\gamma \lesssim 3.1\times 10^{-7}$ for $50$ e-folds and $\gamma\lesssim 2.1\times 10^{-7}$ for 60 e-folds, and again, graphically the allowed region is the one ranging from the turn around point towards the $\gamma \to 0$ limit. By lowering the $x$ ratio higher values of $\gamma$ are admissible and thus higher graphical area. For the Higgs potential, by setting $x=5$, we obtain the following constraints: $\gamma \lesssim 0.0049$ for $N_e=50$ and $\gamma \lesssim 0.0041$ for $N_e=60$. For $x=2$, we restrict $\gamma \lesssim 0.04$ for the range $N_e \in [50,60]$.

After inflation, the pressure is kinetically dominated $p\approx K$, as discussed above, and $F_1+\frac{F_2\mathcal{L}}{\kappa}$ remains positive throughout the whole history of the Universe. In the latter case, the behaviour should be identical to GR since one is in the low curvature regime, and the typical mass scale of the non-minimal coupling function is of the order of $10^{13}$ GeV as it is noted at the end of Subsection (\ref{subsec:consistency}). Furthermore, $F_1+\frac{F_2\mathcal{L}}{\kappa}<\infty$, because neither the curvature nor the inflaton Lagrangian, i.e. its pressure, blow up. Therefore, the homogeneity and isotropy of the Universe depicted in the chosen metric Eq. (\ref{metric}) are not spoiled.

However, due to the high complexity of the full equations and the reheating model-dependency a singularity could appear between inflation and radiation era. Nevertheless, this singularity may potentially be removed when considering the effects of both matter and tensor perturbations, analogously to other modified gravity inflationary scenarios. In this case the above mentioned constraints will not apply, which motivates a dedicated analysis of these non-trivial issues that we leave for future work.

\vspace{1,5cm}
\noindent
{\it \bf Acknowledgements: } The work of C.\ G. is supported by Fundação para a Ciência e a Tecnologia (FCT) under the grant SFRH/BD/102820/2014. J.\,G.\,R. is supported by the FCT Grant No.~SFRH/BPD/85969/2012 and partially by the H2020-MSCA-RISE-2015 Grant No.~StronGrHEP-690904, and by the CIDMA Project No.~UID/MAT/04106/2013.

%\end{acknowledgements}

\end{document}